\documentclass[%
 reprint,
 bibnotes,
 amsmath,amssymb,
 aps,
]{revtex4-2}
\usepackage{graphicx} 
\usepackage{siunitx} 
\usepackage{enumerate}
\usepackage[caption=false]{subfig}
\usepackage{comment}

\begin{document}

\title{The nonlinear limit of Babinet's Principle}%
\author{Valentin Dichtl}
\affiliation{Experimental Physics III, University of Bayreuth, 95447 Bayreuth, Germany}

\author{Thorsten Schumacher} 
\affiliation{Experimental Physics III, University of Bayreuth, 95447 Bayreuth, Germany}

\author{Markus Lippitz} 
\email{markus.lippitz@uni-bayreuth.de}
\affiliation{Experimental Physics III, University of Bayreuth, 95447 Bayreuth, Germany}

\date{\today} 

\begin{abstract}
Babinet's principle is a powerful tool for predicting the scattering behavior of planar structures where the solution for the complementary structure is already known. This makes it ubiquitous in the design of aperture antennas or metamaterials. Even for plasmonic nanostructures, a qualitative match of the behavior for complementary structures has been reported.
Here, we discuss whether Babinet's principle can be extended to nonlinear scattering. We compare the third harmonic emission of plasmonic nanorods and complementary nanoslits by far field imaging and simulation. We find significantly different far field images, in agreement between experiment and simulation. We explain these differences by the higher spatial resolution at the third harmonic wavelength and by additional eddy currents in slits that are not present in rods.
Within these limits, Babinet's principle can guide the design of inverted nonlinear plasmonic resonators, which promise to be more stable at high excitation power due to better thermal conductivity.
\end{abstract}

\maketitle 

\section{Introduction}

\begin{figure}
        \includegraphics[]{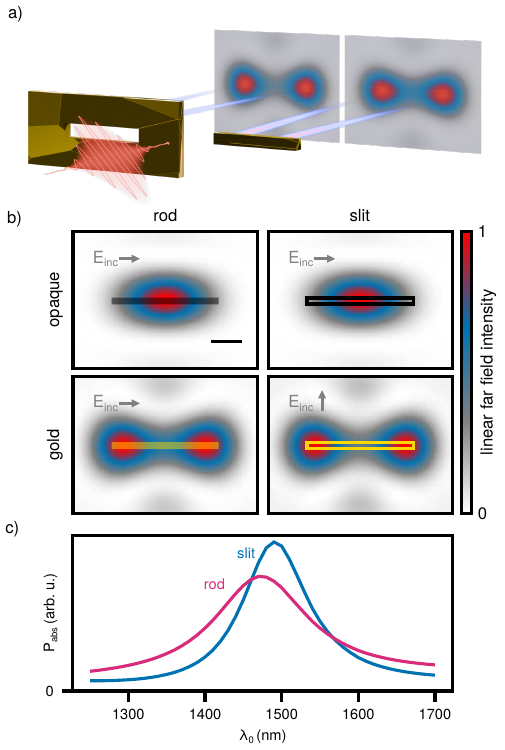}  
        \caption{\textbf{Babinet's principle on the nanoscale.} 
        a) The far field images of a plasmonic slit and a rod are very similar when the linear polarization direction is rotated by 90 degrees.
        b) While Babinet's principle holds exact for ideal opaque materials, it is also approximately valid for plasmonic nanostructures, as shown here for far field images of a nanorod and a nanoslit superimposed with a sketch of the object (scale bar: \SI{500}{\nm}).
        c) The same applies to the absorption spectrum, where the absorption of a slit is defined as the difference in absorption of a plane due to the presence of a slit.  The resonance is the third-order plasmonic eigenmode at $\lambda_\text{plasmon} \approx 2/3\, L_\text{structure}$.
        }
        \label{fig:Babinet_lin}%
\end{figure}

Jacques Babinet formulated his famous principle for diffraction at opaque materials: The image produced by placing a small opaque disc in a beam of light is identical to that of a screen, in which everything is opaque except an aperture the size of the former disk \cite{Born.2019,Pelosi.2017}.
A simulation of Babinet's principle for an opaque rod and slit is sketched in Fig.~\ref{fig:Babinet_lin}a.
In modern optics, Babinet's principle is derived using linear field equations and perfect electric conductors \cite{Harrington.2001}. This extension of Babinet's principle was first presented by H.\,G.\,Booker in 1946 \cite{Booker.1946}.  In contrast to Babinet's formulation where only the screen is changed to its complementary, one also needs to change the light source to its dual: In the case of a linearly polarized plane wave incident on the conducting screen, one needs to rotate the polarization by ninety degrees \cite{Hand.2008}. Then in both cases the  scattered fields result in the same intensity distribution on an imaging screen, up to some factor. This is well suited  for radio frequency (RF) applications where the penetration of the electric field into the conducting metal can be neglected, and only surface currents contribute \cite{Jackson.1975}.

Regarding metal nanostructures in the field of plasmonics, a qualitative agreement with Babinet's principle  has been found, although the skin depth can no longer be neglected \cite{Dorfmuller.2010,Liu.2022,Horak.2019,Zentgraf.2007,Yang.2014,Huck.2015,Feth.2008}.  The term 'quasi-Babinet principle' was coined \cite{Hamidi.2025}. 
Plasmonic nanorods and nanoslits show very similar far field images of their plasmonic hot spots  and resonance spectra (Fig.~\ref{fig:Babinet_lin}b, c). 
Microscopically, this is explained by replacing electric by magnetic dipoles and vice versa \cite{Hentschel.2013,Yang.2014,Hand.2008,Falcone.2004}.

Plasmonic nanostructures differ from many other materials by their strong nonlinear effects \cite{Obermeier2018}. The huge volume nonlinearity of plasmonic metals leads to third-harmonic generation (THG) in plasmonic hot spots. In the near field, linear and nonlinear fields are very similar \cite{Obermeier2018}, as one is the source of the other:
\begin{equation}
    P_{3\omega}(\boldsymbol{r}) = \epsilon_0 \,\chi^{(3)} \, E_{\omega}(\boldsymbol{r})^3  \quad . 
\end{equation}
The linear far field follows Babinet's principle. Does so also the nonlinear far field? Here, we  explore to which extent  Babinet's principle can be applied in the nonlinear case.

\section{Fixed Fundamental Wavelength}

As can be seen in Fig.~\ref{fig:Babinet_lin}c, the linear response of fully complementary structures is spectrally slightly different. In addition, it is difficult if not impossible to fabricate exact complementary structures. In order to avoid strong deviations by a slight resonance mismatch, we chose wavelengths that are blue-shifted by about half a resonance width with respect to the third-order plasmonic eigenmode $\lambda_\text{plasmon} \approx 2/3 \, L_\text{structure}$. As we will see below, the spectral variation of the far field image is weak in this spectral region.

\begin{figure}
    \includegraphics[]{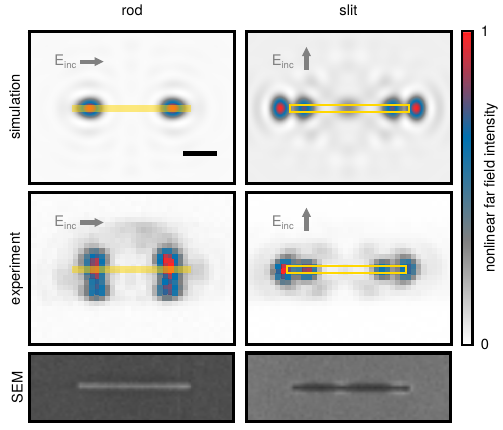} 
    
    \caption{\textbf{Third-harmonic emission far field images.}
    Simulated and measured nonlinear far field image for a rod (left, \(\lambda_\text{0,rod} = \SI{1375}{nm}\)) and a slit  (right, \(\lambda_\text{0,slit} = \SI{1425}{nm}\)). 
    To emphasize the visibility of the hotspots, a nonlinear color scale is used. The dimensions of the structures are depicted by the golden insets (scale bar: \SI{500}{nm}). The `stretching' in y-direction of hotspots measured for the rod probably stem from gold residues, since this is not observed at samples manufactured with electron beam lithography.
    }\label{fig:Babinet_nonlin}
\end{figure}

Let us start by comparing  third-harmonic (TH) far field emission images of a complementary plasmonic nanorod and nanoslit at a fixed fundamental wavelength (Figure~\ref{fig:Babinet_nonlin}).  While experiment and simulation agree reasonably well, the complementary structures show significant differences in their third-harmonic far field image. The rod shows TH emission in the outer quarter of its length. In contrast, the slit emission is centered around the ends of the slits and includes a node at the slit ends that divides each spot into two parts. Changing the structure to its complementary thus shifts the emission centers and changes the shape of the hot spot. Thus, Babinet's principle holds only in a rather weak form in the nonlinear case.

\section{Microscopic Explanations} 

A crucial aspect of Babinet's principle is the imaging of the scattered light onto a screen in the far field. From the perspective of Fourier optics, this propagation in the far field makes the images identical, even though the near fields of the complementary structure are different.

The near field in the object plane can be decomposed into plane waves with different in-plane wave vector $k_\parallel$, which is identical to the two-dimensional Fourier transform. But only plane waves with $|k_\parallel | \le  k_0 = 2\pi / \lambda $ can contribute to the far field image in free space \cite{Goodman.1996,Novotny.2012}.  The Fourier transform of the near fields must be identical inside a circle of radius \(k_0\) in order to look identical and thus obey Babinet's principle.

\begin{figure}
    \includegraphics[]{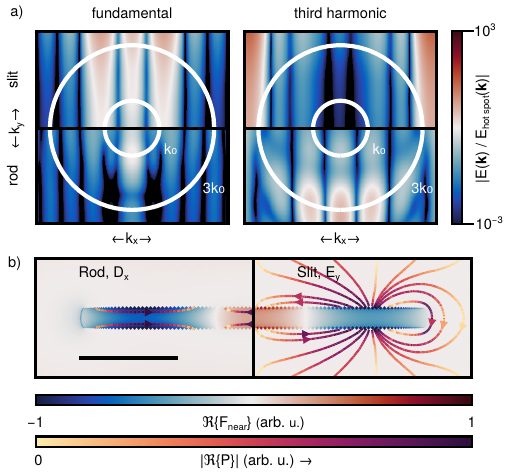}
    \caption{\textbf{Near field and plasmonic hotspots.}
    a) Amplitude of the linear and nonlinear near field components for rod and slit in Fourier space on a logarithmic clipped  colorscale. 
    A  point spread function representing the typical hotspot size (\SI{100}{nm}) is normalized out. The power of the fundamental fields inside the circle $ k \le k_0$ is identical.
    b)  The dual field components \(F_\text{near}\) displacement field $D_x$ and electrical field $E_y$ are mainly responsible for THG in rod and slit, respectively. The streams depict the polarization density $P$, which is  proportional to the current density.
   }\label{fig:K-Space}
\end{figure}

This circle is shown in Figure~\ref{fig:K-Space}a together with the Fourier transform of the linear and nonlinear near field for a gold rod and slit. To emphasize the effect of the structure, we remove the influence of the  shape of the hot spots by dividing in Fourier space by a Gaussian of the reciprocal hot-spot size of \SI{100}{nm}. 

As expected from the plasmonic quasi-Babinet principle, the Fourier components within the circle of radius \(k_0\) are very similar for slit and rod at the fundamental wavelength.
This similarity ends for larger wave vectors, but these fields do not propagate in the far field.
This is different for imaging the third harmonic, because here the far field can contain information up to a wave vector of \(3k_0\).
In addition, third harmonic generation is a local process, and all Fourier components enter. The nonlinear near field and its Fourier transform (right panel of Fig.~\ref{fig:K-Space}a) are thus markedly different for rod and slit, regardless of the length of the wave vector.

From the perspective of Fourier optics, Babinet's principle breaks down for third-harmonic emission because the shorter third-harmonic wavelength contains a 'super-resolution' effect that reveals differences in the fundamental near fields that would otherwise not propagate to the far field.

The polarization (or current) density (Fig.~\ref{fig:K-Space}b) gives further insight.
Inside the rod, a cosine-shaped polarization density forms along the long axis of the rod. It is mostly aligned with the electric field in the x-direction, which is also the polarization direction of the TH emission in the far field.

This is different for the slit. Here, these currents have an eddy current like shape, whose resulting magnetic field can be related to a magnetic dipole moment perpendicular to the sample plane \cite{Ogut.2011,Hentschel.2013}.
The current loops are broken by the slit, creating jamming points. 
This leads to cosine-shaped hotspots and field enhancement along the long slit axis, similar to the rod.  
Unlike the rod, the orientation of these fields is perpendicular to the long axis, as  the polarization components oriented along the x-direction interfere destructively in the far field.

The large current loops around the ends of the slit have no equivalent in the rod. They lead to additional nonlinear emission that is not present in the dual counterpart. From the perspective of microscopic current density, Babinet's principle breaks down for third-harmonic emission due to the additional eddy currents around the ends of the slit. In linear emission, they cannot be spatially resolved and merge with the hot spots along the structure. In nonlinear emission, they can be clearly observed.

\section{Tuning the  Fundamental Wavelength}

Babinet's principle is usually discussed for a single wavelength.
For plasmonic nanostructures, the particle resonances come into play and their phase relation depends on the chosen wavelength. We have exploited this effect in previous work \cite{Wolf.2016} to switch the nonlinear near field of a plasmonic nanorod.

\begin{figure}
    \includegraphics[]{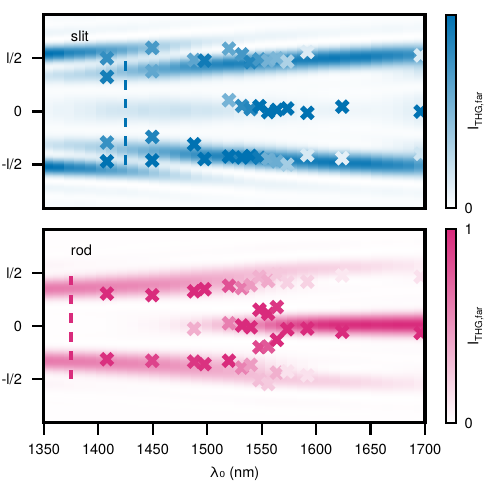}
    \caption{%
    \textbf{Wavelength-dependent far field emission pattern.} The third-harmonic emission pattern differs significantly between  rod (bottom) and  slit (top) and thus do not follow Babinet's principle, especially in the red spectral range. The panels show a line cut through the pattern as a function of the fundamental wavelength and compares the simulation (heat map) with the experimental peak positions (symbols). 
    The extent of each structure is marked with \(l/2\). Dashed lines of length \(l\) display the wavelengths at which the comparisons in Fig.\,\ref{fig:Babinet_lin} and Fig.\,\ref{fig:Babinet_nonlin} are done.
    Experiment and simulation are in good agreement. For the slit in the red spectral region, amplitudes differ from the simulation. 
    Since the measurement is performed with the structures placed on glass coverslips with a refractive index of \num{1.5}, we divide the measurement wavelength by an effective index of refraction of \num{1.25} for better comparability.
    }\label{fig:Transition}
\end{figure}

A nanorod has multiple Fabry-Perot like resonances, at $\lambda_\text{plasmon} \approx. 2/n \, L_\text{structure}$ with integer order $n$. Even orders are dark for perpendicular plane wave excitation. The phase relation between the modal fields changes when the fundamental wavelength is tuned over, e.g.,  the third-order resonance. Blue of the resonance, fundamental and third-order modes are in phase, red of the resonance they are 180 degrees out of phase.
The \(\arctan\) shape of the phase curve `amplifies' the influence of small frequency changes. 
As the modes add up including their phases, constructive and destructive interference can be interchanged, as the fundamental frequency is tuned through a single resonance. The effect is amplified and becomes clearly visible in the case of nonlinear emission \cite{Wolf.2016}.
In contrast to  our former work \cite{Wolf.2016}, here we do not use back focal plane imaging, but image the real space.
Nonetheless, we find both in simulation and measurement the characteristic double to single spot switching for a plasmonic nanorod (bottom panel of Fig.\,\ref{fig:Transition}).

The switching behavior for slits on the other hand differs.
The simple Fabry Perot model does not consider the additional hotspots forming at the ends of the slit structure.
As they are inherent to every plasmonic eigenmode they will always alter the nonlinear far field of a slit compared to that of a rod.
The top panel in Fig.\,\ref{fig:Transition} shows that at resonance the emission pattern can be described by a four-spot pattern where each hotspot is positioned at the outer quarters of the structure. 
Moving further to the red part of the spectrum, the outer spots vanish and are surpassed by the former inner spots.   
This is due to the interference of the fields of the fundamental and third eigenmode, which interfere constructively when out of phase and destructively when in phase at the short edges of the slit.

In contrast to the slit, for the rod the relative amplitudes of the peaks extracted from the measurements do not agree very well with the simulations for the red part of the spectrum.  
In the experiment, the amplitude of the middle hotspot dominates over the outer ones, which is the opposite of the simulation results.  
We assume that this is caused by the plane wave excitation in the simulations in contrast to the finite focus size in the experiment, which is almost equal to the structure size. Since the gold layer extends well beyond the laser focus, the system may be more sensitive to deviations from plane wave excitation than the rod antenna.

\section{Conclusion} 

Babinet's principle is a powerful tool for estimating the optical response of dual structures. It holds reasonably well for the fundamental field to guide the design of plasmonic nanostructures. We have shown through experiments and numerical simulations that one has to be careful when applying Babinet's principle to nonlinear hot spots. Here, the differences between dual structures are much larger than in the linear case. We explain this observation from two perspectives: In Fourier optics, the reduced wavelength of the third harmonic leads to a “super-resolution” effect that makes  differences visible that remain hidden for the fundamental light. The microscopic current density explains the origin of these differences: the eddy currents along the ends of the slit lead to additional hot spots that are only discernible at the third-harmonic wavelength.

Due to the large metal plane, plasmonic slits offer better thermal cooling, lower lattice temperature and thus higher stability at resonant excitation than plasmonic rods \cite{Melentiev2016,Melentiev.2017,Huck.2015},  making them promising candidates for nonlinear plasmonics. Our work outlines the nonlinear limits of Babinet's principle and thus helps to design inverted nonlinear plasmonic resonators.

\section*{Methods} 

\subsection{Experimental setup and sample preparation}

Our nanostructures are fabricated by Focused Ion Beam milling (FIB) from a monocrystalline gold flake with a thickness of \SI{100}{nm}. 
We followed the recipe of Ref.~\cite{Krauss.2018} for growing and depositing the gold flakes on \SI{170}{\micro\metre} thin coverslips.  
The slit shown has a measured length of \SI{1.8}{\micro\metre}, while the rod has a length of \SI{1.75}{nm}. The width is estimated to be \SI{100}{nm}.

An optical parametric oscillator pumped by a Ti:sapphire laser generates pulses (duration \SI{200}{fs}, repetition rate \SI{76}{MHz}) in the wavelength range \SI{1700}{nm} to \SI{2200}{nm}. 
A Glen-Thompson prism together with a half-wave plate allows adjustment of the linear polarization direction.
An aspherical lens (NA 0.5) produces a focus (FWHM \SI{1800}{nm}) slightly larger than the size of the nanostructures. The sample is positioned by a 
three-axis piezo stage.

The emitted third harmonic light is collected by an oil immersion objective (Olympus UPlanSApo \(60\times\), NA 1.35). 
A filter (\(2\times\)SCHOTT 2 KG 5) removes other mixing products, multi-photon luminescence and the fundamental laser beam.
We image (f = \SI{600}{mm}) the sample plane on the entrance of a spectrometer (IsoPlane SCT320) equipped with an EM CCD camera (excelon ProEM HS \(1024\times1024\)), resulting in an effective pixel size of \SI{77}{nm}. We use semi-spectral images with the first order of diffraction to convince ourselves that the collected light is indeed at the third harmonic. We then switch to the 0th diffraction order to obtain the images shown in Fig.\,\ref{fig:Babinet_nonlin}.

\subsection{Simulation methods} 

In order to compute the linear and nonlinear properties of our nanostructures, we use the commercially available finite element solver Consol Multiphysics in frequency domain. While the dielectric function of gold is based on the data provided by Olmon et al. \cite{Olmon.2012} for single crystalline gold, we waive the glass substrate of the experiment and use an effective medium with a refractive index of \num{1.0}. The gold thickness is matched to our sample with \SI{100}{nm}. The slit as well as the rod have identical dimensions of \SI{1742}{nm} $\times$ \SI{100}{nm}. 

To calculate the linear absorption spectrum and the far field image, we proceed as follows: First, we use the scattering-field method for calculating the scattered field for a certain excitation field without the nanostructure. In case of the rod, this is simply a plane wave. For the slit, we calculate the plane wave excitation field through the homogeneous gold film by an external T--matrix formalism and embed the result analytically into our Comsol model. The superposition of excitation- and scattered field then gives us the overall field at a given frequency \(\omega_0\). From these field components we  extract the absorption and scattering spectra as well as the complex field components at the shadow side of the structure. Each of these field components gets far field transformed before the time averaged Poynting vector is calculated in the screen plane. The normal component of it with respect to the screen is the far field intensity \cite{Goodman.1996}.

After simulating the linear properties, the material polarization values are raised to the third  power and used as boundary conditions to then solve this system again for a frequency at \( 3 \omega_0 \). The resulting nonlinear spectra and complex fields are analyzed in the same way as for the linear model.  For both steps, we use the biconjugate gradient stabilized method of Comsol.

\bibliography{lit}
\end{document}